\documentclass[final,12pt]{elsarticle}
\usepackage{amssymb}
\usepackage{amsmath,epsfig}
\usepackage[latin1]{inputenc}
\usepackage{graphicx}
\usepackage[english]{babel}
\usepackage{xspace}
\usepackage{subfigure}
\usepackage{float}
\usepackage{multirow}
\usepackage{placeins}
\usepackage[percent]{overpic}
\usepackage{eurosym}
\usepackage{tablefootnote}

\usepackage{array}
\newcolumntype{L}[1]{>{\raggedright\let\newline\\\arraybackslash\hspace{0pt}}m{#1}}
\newcolumntype{C}[1]{>{\centering\let\newline\\\arraybackslash\hspace{0pt}}m{#1}}
\newcolumntype{R}[1]{>{\raggedleft\let\newline\\\arraybackslash\hspace{0pt}}m{#1}}

\usepackage{lineno}

\journal{Irfu Internal note}

\begin{document}

\begin{frontmatter}

%% Title, authors and addresses

\title{What is the theoretical time precision achievable using a dCFD algorithm ?}

\author{Eric Delagnes\corref{cor1}}
\ead{eric.delagnes@cea.fr}

\address{IRFU, CEA, Universit\'{e} Paris-Saclay, F-91191 Gif-sur-Yvette, France}

\cortext[cor1]{Corresponding author}

\begin{abstract}
%% Text of abstract
The time precision achievable using standard analog methods is well known. Several expressions for timing methods using digitized signals have been recently proposed  in workshops, conferences or training courses. Most of them are only partially exact. This paper presents a  comprehensive calculation of the timing precision for algorithms using digital treatment of digitized signal to mimic analog discriminator-based methods. The results of these calculations are discussed for various cases of correlation between samples.  

\end{abstract}

\begin{keyword}
Time picking methods \sep Waveform sampling \sep CFD \sep dCFD \sep SCA \sep TDC
%% keywords here, in the form: keyword \sep keyword

%% MSC codes here, in the form: \MSC code \sep code
%% or \MSC[2008] code \sep code (2000 is the default)

\end{keyword}

\end{frontmatter}

%%
%% Start line numbering here if you want
%%
%%\linenumbers

%% main text
\section{Introduction}
\label{S:1}
	Time picking of pulses with high precision of the order of few ps is a key challenge for the next generation high energy physics experiments. Since decades, precise timing is usually obtained using the combination of an analog discriminator followed by a device measuring the time of arrival of its output and digitizing it (Time to Digital converter or Time to Amplitude converter followed by an \emph{ADC}). The discriminator can be a simple leading edge one or a more sophisticated one. One major effect affecting the precise timing measurement is the time walk, \emph{ie} the dependency of the timing with the input pulse amplitude, which can be partially canceled if a Constant Fraction Discriminator structure \cite{GEDCKE1967377} is used or corrected after calibration using the pulse's amplitude or Time Over Threshold information \cite{SFE16}. Continuous progress of commercial Analog to Digital converter up to few 100 MSPS and the availability of second \cite{Wavecatcher}, \cite{DRS} and third generation analog memories \cite{Sampic} now allow, for a moderate cost, to sample and digitize directly the analog signal at high frequency. The digitized samples can then be digitally processed to extract the pulse's timing. Among the possible algorithm, the digital CFD \emph{dCFD}, which mimics the analog\emph{CFD} behavior is very attractive because it requires only moderate computing resources for performance often similar to those obtained with more complex algorithms \cite{BretonMCPPMT}. If the theoretical expression of the time precision obtained using an analog method is well known, the one used for \emph{dCFD} algorithm is often under or over-estimated. An exact calculation of it is presented in this paper, taking into account the correlation effect between samples.

\section{Time resolution of the leading edge discriminator}
\label{S:2}
\subsection{Analog leading edge discriminator, relationship between voltage noise and jitter}
\begin{figure}[htb]
\centering\includegraphics[width=0.8\linewidth]{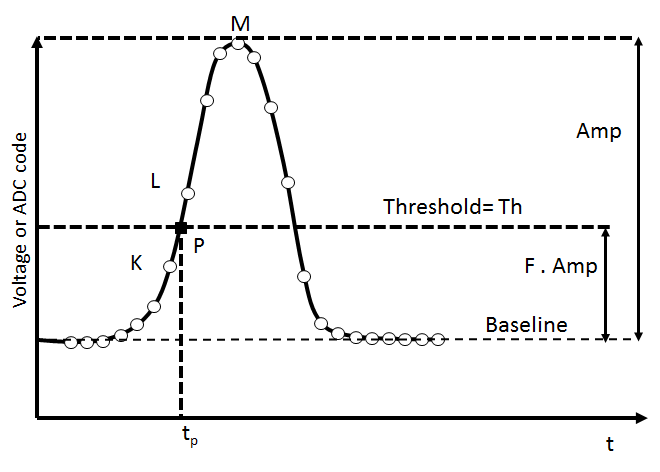}
\caption{Principle of LED, dLED,CFD and dCFD}
\label{Fig1}
\end{figure}

	The simpler time picking method, the Leading Edge Discrimination (\emph{LED}), illustrated by Figure~\ref{Fig1}, consists in measuring the (\emph{t$_{\text{p}}$}) time of the point \emph{P}  at which the analog input signal crosses a fix threshold (\emph{Th}). 

\begin{figure}[htb]
\centering\includegraphics[width=0.8\linewidth]{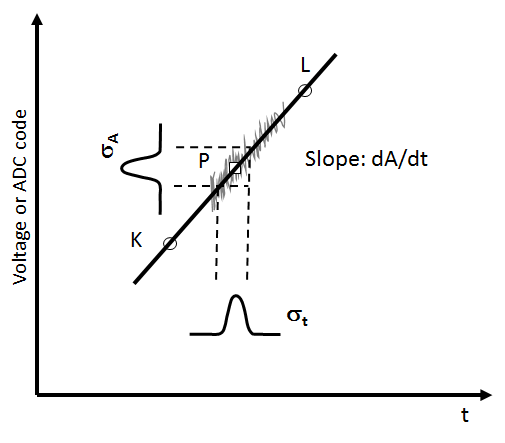}
\caption{Voltage noise and jitter equivalence}
\label{Fig2}
\end{figure}

    As shown on Figure~\ref{Fig2}, if the signal is noisy with a noise standard deviation $\sigma_{\text{A}}$ and if we consider that the signal is linear around \emph{P} with a \emph{dA/dt} slope, we can express $\sigma_{\text{t}}$, the standard deviation of the threshold crossing time (jitter) thanks Equation~\ref{eq:em0}:

\begin{equation}
\label{eq:em0}
\sigma_t = \frac{\sigma_A}{dA/dt} 
\end{equation}

Conversely, if the time of P is measured with a limited precision ($\sigma_{\text{j}}$)- by a TDC for instance - Equation~\ref{eq:em0}, allows us to translate it in an equivalent voltage noise.
In both cases, we have to remember that this equation can only be used if the local linear approximation of the signal remains valid for the amplitude of the noise.

As the TDC jitter and the voltage noise are normally uncorrelated, we can express the total standard variance on the time by  Equation~\ref{eq:em01}

\begin{equation}
\label{eq:em01}
\sigma_{tt}^2 = \sigma_j^2 + (\frac{\sigma_A}{dA/dt)})^2
\end{equation}

This two equations are only valid for an input signal with a fix amplitude. If the amplitude varies, the timing is affected by the time walk effect that is not taken into account here.

\subsection{Digital leading edge discriminator, effect of interpolation}

	If the analog input signal is digitized at a \emph{F} rate (corresponding to a \emph{T}  sampling period), it is possible, as shown on Figure~\ref{Fig1}, to digitally mimic the \emph{LED} method by interpolating between the two samples \emph{K} and \emph{L} around the threshold crossing to find the crossing point \emph{P}. With this Digital Leading Edge Discrimination method (\emph{dLED}), as we use at least two samples to find the timing of the crossing, it is natural to think that the timing resolution of a noisy signal can be improved compared to the \emph{LED} case.
    In this subsection, we will evaluate this time resolution improvement assuming that the signal is linear between the  \emph{K} an \emph{L} samples.
    
\begin{figure}[htb]
\centering\includegraphics[width=0.8\linewidth]{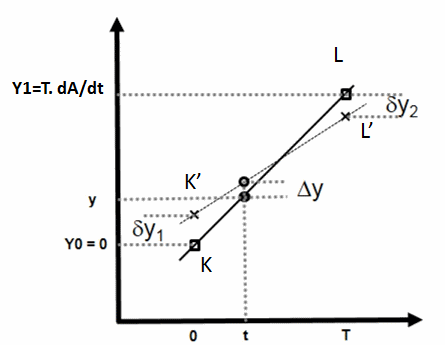}
\caption{Interpolation between two noisy samples}
\label{Fig3}
\end{figure}
    
As shown on Figure~\ref{Fig3}, for simplicity, the coordinates of\emph{K} are (0,0). Those of \emph{L} are (T, TdA/dt) where \emph{dA/dt} represents the slope of the signal and \emph{T} the sampling period. If we consider now consider that the input signal is noisy, \emph{K} and \emph{L}, respectively affected by  \emph{$\delta y1$}  and \emph{$\delta y2$} noise voltages, become \emph{K'} and \emph{L'}.Using \emph{K'} and \emph{L'} to interpolate the values at the time \emph{t} leads to an error \emph{$\Delta$Y} compared to the voltage for the same time on the original segment (K,L). 
\begin{equation}
\label{eq:em1}
\Delta Y = \partial y2\frac{t}{T} +\partial y1\frac{T-t}{T}
\end{equation}

The variance of this error is given by :
\begin{equation}
\label{eq:em2}
Var(\Delta Y)= \frac{\left ( T-t \right )^{2}\,var(\partial y1)+t^{2}\,var(\partial y2)+2\,t\,(T-t)\,cov\left ( \partial y1,\partial y2 \right )}{T^{2}}
\end{equation}

Integrating Equation~\ref{eq:em2} over a sampling period allows to calculate the average error introduced by the noise.

\begin{equation}
\label{eq:em3}
\overline{Var(Y)}=\frac{1}{T}\int_{0}^{T}Var(Y)\: dt
\end{equation}

\begin{equation}
\label{eq:em4}
\overline{Var(Y)}=\frac{1}{T^{3}}\left \{  var(\partial y1) \left [ T^2 t+ \frac{t^{3}}{3}-Tt^2 \right ]_{0}^{T} + var(\partial y2) \left [  \frac{t^{3}}{3}\right ]_{0}^{T}+cov\left(\partial y1,\partial y2 \right ) \left [Tt^2 -2 \frac{t^{3}}{3} \right ]_{0}^{T} \right \}
\end{equation}

\begin{equation}
\label{eq:em5}
\overline{Var(Y)}=  \frac{1}{3}\, var(\partial y1) +\: \frac{1}{3}\, var(\partial y2)  + \frac{1}{3}\, \:cov\left ( \partial y1,\partial y2 \right ) 
\end{equation}

that can be reduced to Equation~\ref{eq:em6} if we make the hypothesis of noise ergodicity so that the noise variance $\sigma_n^2$ is the same for the two consecutive samples

\begin{equation}
\label{eq:em6}
\overline{Var(Y)} = \frac{2}{3} \,{\sigma_{n}}^{2} + \frac{1}{3}\, \:cov\left ( \partial y1,\partial y2 \right )
\end{equation}

If the noise standard deviation is negligible compared to  $\Delta$Y \footnote{so that dA/dt is not affected.},this equation can be used to calculate the variance of the timing precision by applying Equation~\ref{eq:em0}, leading to : 

\begin{equation}
\label{eq:em7}
\overline{Var(t_{p,n})} =\frac{ \frac{2}{3} \,{\sigma_{n}}^{2} + \frac{1}{3}\, \:cov\left ( \partial y1,\partial y2 \right )}{(dA/dt)^2}
\end{equation}

As in the previous section, we can also consider that the \emph{K} and \emph{L} samples are affected by uncertainties due to jitter, respectively \emph{$\delta$t1} and \emph{$\delta$t2}, with a common value of standard deviation \emph{$\sigma_{\text{j}}$}. If we make the assumption that \emph{$\sigma_{\text{j}}$} is negligible compared to \emph{T}  \footnote{again so that dA/dt is not affected.}, the contribution of these jitters to the variance of the interpolated crossing time can be calculated using successively 1. the reversed form of Equation~\ref{eq:em0},2. Equation~\ref{eq:em7} and finally 3. Equation~\ref{eq:em0}, leading to

\begin{equation}
\label{eq:em8}
\overline{Var(t_{p,j})} =\frac{2}{3} \,{\sigma_{j}}^{2} + \frac{1}{3}\, \:cov\left ( \partial t1,\partial t2 \right )
\end{equation}

In the general case, the noise voltage and sampling jitter are uncorrelated so that the total variance of the interpolated crossing time can be obtained by summing the right terms of Equations~\ref{eq:em7} and ~\ref{eq:em8}.

\begin{equation}
\label{eq:em9}
\overline{Var(t_{p})} = \overline{Var(t_{p,n})} + \overline{Var(t_{p,j})}
\end{equation}

\subsection{Interpolation: various cases of correlation between samples}

The results of Equations~\ref{eq:em7} and \ref{eq:em8} are obviously dependent of the correlation between samples.

\subsubsection{Case of uncorrelated samples}

If the fluctuations in amplitude of the two samples are totally uncorrelated, which is the case for: 
\begin{itemize}
\item an unfiltered white noise (high frequency noise compared to the sampling frequency),
\item the part of noise associated to the sampling, quantization and digitizing processes,
\end{itemize}

Equation~\ref{eq:em7} can be reduced to:
\begin{equation}
\label{eq:em11}
\overline{Var(t_{p,n})} = \frac{2}{3}\,\frac{{\sigma_{n}}^{2} }{{(dA/dt)^2}}
\end{equation}

Similarly, if the fluctuation of time sampling are uncorrelated between samples, which can be the case for example for a properly generated clock of an \emph{ADC}, Equation~\ref{eq:em8} can be reduced to:
\begin{equation}
\label{eq:em12}
\overline{Var(t_{p,j})} =\frac{2}{3}\,{\sigma_{j}}^{2} 
\end{equation}

Both equations are showing that the use of the two uncorrelated noisy samples permits to improve (through interpolation) the time picking resolution by a factor $\sqrt(3/2)$ = 1.23 compared to the analog method.
Note that it is significantly less than the $\sqrt2$ factor often used.

\subsubsection{Case of totally correlated samples}

If the voltage noise is totally correlated between the two samples as it is the case for low-pass-filtered-white noise over-sampled:

Equation~\ref{eq:em7} becomes:
\begin{equation}
\label{eq:em14}
\overline{Var(t_{p,n})} = \frac{{\sigma_{n}}^{2} }{{(dA/dt)^2}}
\end{equation}

Similarly if the fluctuation of time sampling are  totally correlated between samples, which can be partially the case in the SCA in which a single master clock is successively delayed to obtain the consecutive sampling commands, Equation~\ref{eq:em8} becomes:
\begin{equation}
\label{eq:em15}
\overline{Var(t_{p,j})} ={\sigma_{j}}^{2} 
\end{equation}

In this case there is no improvement compared to the analog method.

\subsubsection{Case of anti-correlated samples}

If the covariance term between samples is negative (anti-correlated fluctuation between consecutive samples) the improvement factor, compared to the analog method, can be as high as $\sqrt3$ in the case of fully anti-correlated samples\footnote{conversely to all the other tested cases, this value is not confirmed by Monte-Carlo simulations that give an improvement factor of 2} . But this case seems to be practically very unlikely. 

\subsubsection{Case of partially correlated samples}

As an example, let us examine the case of a voltage white noise filtered by low pass filter with a $\tau$ time constant. The covariance term of Equation~\ref{eq:em7} is derived from the noise autocorrelation function calculated in \cite{autocor} giving :

\begin{equation}
\label{eq:em17}
\overline{Var(t_{p,n})} =\frac{ \frac{2}{3} \,{\sigma_{n}}^{2} + \frac{1}{3}\,{\sigma_{n}}^{2}exp(-T/\tau)}{(dA/dt)^2}
\end{equation}

\begin{figure}[htb]
\centering\includegraphics[width=0.8\linewidth]{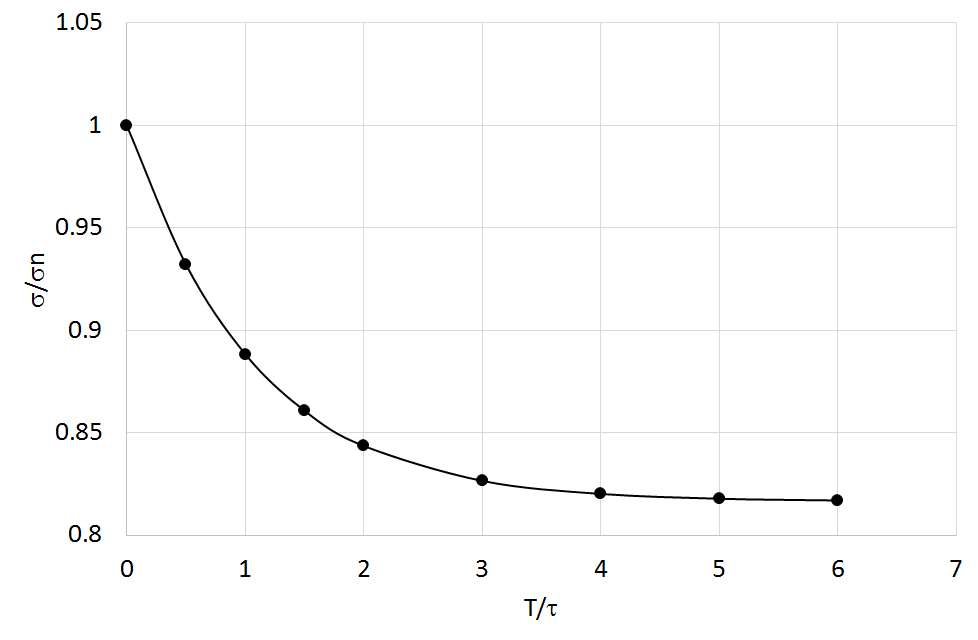}
\caption{Dependency of the noise at interpolated crossing point vs T/$\tau$ in the case of a white noise filtered using a lowpass filter with time $\tau$ time constant }
\label{Fig4}
\end{figure}

The dependency of the inverse of the improvement factor on the standard deviation of the interpolated amplitude at the crossing point  (identical to the one of the interpolated crossing timing)  with respect to the one obtained with the analog method is plotted on Figure~\ref{Fig4} as function of the T/$\tau$ ratio.
If we consider the case of a pulse (output of a charge sensitive amplifier) filtered by a single pole lowpass filter,its rise time is given by 2.2 $\tau$. To extract its timing we need more than 2 samples on the signal rising edge. It means that the T/$\tau$ ratio is smaller that 1 so that the samples used for the interpolation are strongly correlated and that the noise improvement factor will be of only 10$\%$ or less.

\subsection{Digital CFD}

The \emph{dCFD} algorithm, also illustrated by Figure~\ref{Fig1} mimics the behavior of a constant fraction discriminator. For this purpose the amplitude \emph{Amp} of the digitized pulse is determined using a digital algorithm (for example parabolic interpolation around the sample M). A threshold is set to a fraction \emph{F$\times$Amp} of this amplitude, and the signal timing is determined by the crossing point of the interpolated digitized signal and this threshold. As for analog CFD, this method is used to reduce the time walk effect.

The variance of the interpolated timing can be derived from Equation~\ref{eq:em7} by adding an extra  contribution corresponding to the threshold fluctuation equal to the fraction \emph{F} of those of the peak amplitude. Because of the filtering effect of the digital treatment allowing its extraction, the standard deviation of the amplitude is $\alpha\times\sigma_{\text{n}}$ with $\alpha\leqslant1$. If we neglect the noise correlation between the threshold crossing and the peak \footnote{this correlation obviously exists but is much more smaller than between two consecutive samples. As it is likely positive, it will improve very slightly the variance given by Equation~\ref{eq:em18}}, we can calculate the variance of the interpolated crossing time for dCFD as: 

\begin{equation}
\label{eq:em18}
\overline{Var(t_{p})} =\frac{ \frac{2}{3} \,{(1+F^2\alpha^2)\sigma_{n}}^{2}  + \frac{1}{3}\, \:cov\left ( \partial y1,\partial y2 \right )}{(dA/dt)^2} + \frac{2}{3} \,{\sigma_{j}}^{2} + \frac{1}{3}\, \:cov\left ( \partial t1,\partial t2 \right )
\end{equation}

\section{Practical case of a \emph{SCA}}

	Now let us study the case of a typical SCA-based digitizer as SAMPIC. The "voltage" noise digitized by the chain is the quadratic sum of :
\begin{itemize}
\item a voltage noise, with $\sigma_{\text{n}}$ standard deviation and a correlation between consecutive samples depending on the noise spectrum quantified by their covariance (here noted as cov(T))
\item the sampling, quantization and digitization noise, with $\sigma_{\text{d}}$ standard deviation uncorrelated from sample to sample
\end{itemize}

The time sampling precision is the combination of:
\begin{itemize}
\item the jitter on the master clock, with $\sigma_{\text{ck}}$ standard deviation, introducing a correlation on the timing of the N consecutive delays of the DLL providing the sampling signal. The covariance term quantifying the correlation between the two consecutive samples is $\sigma_{\text{ck}}^2{(N-1)/N} \sim\sigma_{\text{ck}}^2 $ 

\item the jitter $\sigma_{\text{step}}$ introduced by the delay steps, which can be considered as uncorrelated from sample to sample 
\end{itemize}

Equation~\ref{eq:em18} then becomes :
\begin{equation}
\label{eq:em19}
\overline{Var(t_{p})} =\frac{ \frac{2}{3} \,{(1+F^2\alpha^2)({\sigma_{n}}^{2}+{\sigma_{d}}^{2})  + \frac{1}{3}\, \:cov\left ( T \right )}}{(dA/dt)^2} + \frac{2}{3} \,{\sigma_{step}}^{2} +  {\sigma_{ck}}^{2} 
\end{equation}

 \section{Conclusion}  
 
	Theoretical expression for the timing resolution achievable using \emph{dLED} and \emph{dCFD} with linear interpolation methods have been calculated assuming:
   
\begin{itemize}
\item A linear signal between samples,
\item Voltage noises and time jitters small compared to respectively the voltage difference between the 2 interesting consecutive samples and the sampling period.  
\end{itemize}

It has been shown that these methods permits to improve the timing by only up to 20$\%$ compared to the basic analog method. This improvement decreases when the consecutive samples are correlated.

Even if only the linear interpolation and a piecewise linear signal have been studied, the trends identified in this paper will remain valid in a more general case.   

\section{Acknowledgments}
This work has been partially funded by the P2IO LabEx (ANR-10-LABX-0038) in the framework "Investissements d'Avenir" (ANR-11-IDEX-0003-01) managed by the French National Research Agency (ANR).

%%\clearpage

%% References with bibTeX database:

%\bibliographystyle{unsrt}
\bibliography{biblio}

\begin{thebibliography}{1}

\bibitem{GEDCKE1967377}
D.A. Gedcke and W.J. McDonald.
\newblock A constant fraction of pulse height trigger for optimum time
  resolution.
\newblock {\em Nuclear Instruments and Methods}, 55:377 -- 380, 1967.

\bibitem{SFE16}
E.~Delagnes, P.~Abbon, Y.~Bedfer, J.~C. Faivre, F.~Kunne, A.~Magnon,
  S.~Platchkov, P.~Rebourgerad, and D.~Thers.
\newblock Sfe16, a low noise front-end integrated circuit dedicated to the
  read-out of large micromegas detectors.
\newblock {\em IEEE Transactions on Nuclear Science}, 47(4):1447--1453, Aug
  2000.

\bibitem{Wavecatcher}
D.~Breton, E.~Delagnes, J.~Maalmi, and P.~Rusquart.
\newblock The wavecatcher family of sca-based 12-bit 3.2-gs/s fast digitizers.
\newblock In {\em Real Time Conference (RT), 2014 19th IEEE-NPSS}, pages 1--8,
  May 2014.

\bibitem{DRS}
Stefan Ritt, Roberto Dinapoli, and Ueli Hartmann.
\newblock Application of the \{DRS\} chip for fast waveform digitizing.
\newblock {\em Nuclear Instruments and Methods in Physics Research Section A:
  Accelerators, Spectrometers, Detectors and Associated Equipment}, 623(1):486
  -- 488, 2010.
\newblock 1st International Conference on Technology and Instrumentation in
  Particle Physics.

\bibitem{Sampic}
E.~Delagnes, D.~Breton, H.~Grabas, J.~Maalmi, and P.~Rusquart.
\newblock {Reaching a few picosecond timing precision with the 16-channel
  digitizer and timestamper \textsc{SAMPIC} ASIC}.
\newblock {\em Nuclear Instruments and Methods in Physics Research Section A:
  Accelerators, Spectrometers, Detectors and Associated Equipment}, 787:245 --
  249, 2015.

\bibitem{BretonMCPPMT}
D.~Breton, E.~Delagnes, J.~Maalmi, K.~Nishimura, L.L. Ruckman, G.~Varner, and
  J.~Va'vra.
\newblock High resolution photon timing with mcp-pmts: A comparison of a
  commercial constant fraction discriminator (cfd) with the asic-based waveform
  digitizers \{TARGET\} and wavecatcher.
\newblock {\em Nuclear Instruments and Methods in Physics Research Section A:
  Accelerators, Spectrometers, Detectors and Associated Equipment}, 629(1):123
  -- 132, 2011.

\bibitem{autocor}
J.~Loren Passmore, Brandon~C. Collings, and Peter~J. Collings.
\newblock Autocorrelation of electrical noise: An undergraduate experiment.
\newblock {\em American Journal of Physics}, 63(7):592--595, 1995.

\end{thebibliography}

\end{document}